# Riemannian Geometry Framed as a Generalized Lie Algebra to Incorporate General Relativity with Quantum Theory, I

Joseph E. Johnson[1]

1. Distinguished Professor Emeritus, Department of Physics & Astronomy, University of South Carolina, Columbia SC, 29208 ;
jjohnson@sc.edu  © September 4, 2022 Joseph E. Johnson

**Abstract:** This paper reframes Riemannian geometry (RG) as a generalized Lie algebra allowing the equations of both RG and then General Relativity (GR) to be expressed as commutation relations among fundamental operators. We begin with an Abelian Lie algebra of n operators, X, whose simultaneous eigenvalues, y, define a real n-dimensional space R(n). Then with n new operators defined as independent functions, X'(X), we define contravariant and covariant tensors in terms of their eigenvalues, y', on a Hilbert space representation. We then define n additional operators, D, whose exponential map is to translate X as defined by a noncommutative algebra of operators (observables) where the "structure constants" are shown to be the metric functions of the X operators thus allowing for spatial curvature resulting in a noncommutativity among the D operators. The D operators then have a Hilbert space position-diagonal representation as generalized differential operators plus an arbitrary vector function A(X), which, with the metric, written as a commutator, can express the Christoffel symbols, and the Riemann, Ricci and other tensors as commutators in this representation. Traditional RG and GR are obtained in a position diagonal representation of this noncommutative algebra of 2n+1 operators. Our motivation was suggested by the fact that Quantum Theory (QT), Special Relativity (SR), and the Standard Model (SM) are framed and well-established in terms of Lie algebras. But GR, while also well-established, is framed in terms of nonlinear differential equations for the space-time metric and space-time variables. We seek to provide a more general framework for RG to support an integration of GR, QT, and the SM by generalizing Lie algebras as described. The gauge transformations that here support the SM are altered by the inclusion of the metric. Other non-trivial consequences are discussed. Finally, we initially explore the determination of the asymmetric components of the metric.

(298 words 1994 characters w spaces)

**Keywords:** Riemannian Geometry, Lie algebra, Quantum Theory, metric, Heisenberg algebra, General Relativity, Standard Model, Dark Matter

## 1. Introduction

Lie algebras and the Lie groups which they generate have played a central role in both mathematics and theoretical physics since their introduction by Sophius Lie in 1888 [1]. Both relativistic quantum theory (QT) and the gauge algebras of the phenomenological standard model (SM) of particles and their interactions are framed in terms of observables which form Lie algebras and are firmly established [2,3,4,5]. The Heisenberg Lie algebra (HA) among (generalized) momentum and position operators, [D, X] gives the foundational structure of QT and also has applications in mathematics in studies related to Fourier transforms and harmonic analysis [6,7,8,9]. Likewise, in QT one has the Poincare symmetry Lie algebra (PA) of space-time observables whose representations define free particles.

But the theory of gravitation as expressed in Einstein's general theory of relativity (GR), although also firmly established, is formulated in terms of a Riemannian geometry (RG) of a curved space-time where the metric is determined by nonlinear differential equations from the distribution of matter and energy [10,11]. In GR there are no operators representing observables, and thus no commutation rules to define Lie algebras, and thus no representations of such algebras. The observables in GR are (a) the positions of events in space-time, and (b) the metric function of position in space-time (and its derivatives) which define the distance between events, and which define the curvature of space-time. Thus QT and GR are expressed in totally different mathematical frameworks and their merger into a single theory has been a central problem in physics for over a century.

However, the space-time events in QT are the eigenvalues of the space-time operators which are an essential part of the HA which also contains the Minkowski metric which defines the associated translated distance when space-time is not curved. If the associated space were curved, one would have a metric that was a function of the position in space-time. Such a generalized HA would no longer allow closure as a traditional Lie algebra but rather closure in the enveloping algebra of analytic functions of the basis elements of the Lie algebra.

This led us to consider a Generalized Lie Algebra (GLA) (a noncommutative operator algebra, NOA), generalizing the framework of a Lie algebra, with n space-time operators, $X^\mu$, and n corresponding operators $D^\mu$, which by definition are to execute infinitesimal translations in the associated representation space of the $X^\mu$ eigenvalues, $y^\mu$ ($\mu = 0, 1, .n-1$). The $X^\mu$ are to form an Abelian algebra whose eigenvalues represent a "space-time" manifold of four or a larger number of dimensions as the associated X eigenvalues are simultaneously measurable. But we allow the space to be curved so the corresponding $D^\mu$ operators will not in general commute as the translations in such a curved space can interfere with each other. We found that this approach generalized the HA "structure constants" to be the Riemann metric thus allowing the metric to be a function of the position operators, X, in the algebra [12]. This generalizes the concept of a Lie algebra to allow for "structure constants" that are functions of the X operators in the algebra and thus are no longer constants except approximately in small neighborhoods.

This paper formally reframes RG [13] as a GLA including the HA. We show that the fundamental concepts in RG such as the coordinate transformations, contravariant and covariant tensors, Christoffel symbols, Riemann and Ricci tensors, and the Riemann covariant derivative can all be expressed in terms of commutation relations among these fundamental operators. This framework is reminiscent of contractions of Lie algebras where the structure constants as functions are modified to vary smoothly among different algebras based upon certain external parameters [14, 15, 16, 17, 18, 19] but not in the algebra itself as we propose. In a similar way, our algebra allows the structure constants to be dependent upon the X operators in the algebra so that RG is retrieved as a position-diagonal representation of the algebra as one moves over the Riemann manifold of X eigenvalues.

## 2. Riemannian Geometry Framed as a Noncommutative Algebraic Geometry of Observables

We begin by defining a purely mathematical structure devoid of QT, SM, & GR content. Consider a set of n independent linear self-adjoint operators, $X^\mu$, which form an Abelian Lie algebra of order n, where

$$[X^\mu, X^\nu] = 0 \text{ and where } \mu, \nu = 0, 1, 2, \ldots (n-1). \tag{2.1}$$

Consider a Hilbert space of square integrable complex functions $|\Psi\rangle$ as a representation space for this algebra where a scalar product is used to normalize the vectors to unity: $\langle\Psi|\Psi\rangle = 1$. The simultaneous eigenvectors of the Abelian Lie algebra can be written as the outer product of the $X^\mu$ eigenvectors with the Dirac notation

$$|y^0\rangle |y^1\rangle |y^2\rangle \ldots |y^{n-1}\rangle = |y^0, y^1, y^2, \ldots y^{n-1}\rangle = |y\rangle \tag{2.2}$$

where the eigenvalues $y^\mu$ label the associated eigenvectors $|y\rangle$ of the $X^\mu$ operators and where we use the notation

$$X^\mu |y\rangle = y^\mu |y\rangle \text{ where the } y^\mu \text{ are real numbers.} \tag{2.3}$$

These independent real variables $y^\mu$ can be thought of as the coordinates (or basis vectors) of an n-dimensional space $R_n$ since each set of values defines a point in $R_n$. Let the eigenvectors be normalized to be orthonormal with the scalar product

$$\langle y_a | y_b \rangle = \delta(y^0_a - y^0_b)\, \delta(y^1_a - y^1_b) \ldots \delta(y^{n-1}_a - y^{n-1}_b). \tag{2.4}$$

Let the decomposition of unity

$$1 = \int_{dy} |y\rangle\langle y| \tag{2.5}$$

project the entire space onto the basis vectors $|y\rangle$ where $\langle y|$, using Dirac notation, is the dual vector to $|y\rangle$. A general vector in the representation (Hilbert) space of this Lie algebra can then be written as

$$|\Psi\rangle = \int_{dy} |y\rangle\langle y|\Psi\rangle = \int_{dy} \Psi(y) |y\rangle, \tag{2.6}$$

where the function $\Psi(y)$ gives the "components" of the abstract vector $|\Psi\rangle$ on the basis vectors $|y\rangle$. Thus

$$\langle\Psi|\Psi\rangle = 1 = \int_{dy} \langle\Psi|y\rangle\langle y|\Psi\rangle = \int_{dy} \Psi^*(y)\Psi(y). \tag{2.7}$$

Now consider another set of n linear operators, $X'^\mu$, which are independent analytic functions, $X'^\mu(X^\mu)$, of the $X^\mu$ operators also forming an Abelian Lie algebra on the same representation space for this algebra where it follows that

$$[X'^\mu, X'^\nu] = 0. \tag{2.8}$$

Let the $X'^\mu$ have eigenvectors $|y'\rangle$ and eigenvalues $y'^\mu$ given by

$$X'^\mu |y'\rangle = y'^\mu |y'\rangle \text{ where } y'^\mu \text{ are real numbers.} \tag{2.9}$$

The same orthonormality and decomposition of unity also obtain for the $|y'\rangle$ vectors which are also to be a complete basis for the space. Then we can let the $X'^\mu(X^\nu)$ act to the left on the dual vector $\langle y'|$ and also act to the right on the vector $|y\rangle$ as

$$\langle y' | X'^\mu | y \rangle = \langle y' | X'^\mu(X^\nu) | y \rangle \text{ to give} \tag{2.10}$$

$$y'^\mu \langle y' | y \rangle = y'^\mu(y) \langle y' | y \rangle. \tag{2.11}$$

Thus the eigenvalues $y'^\mu = y'^\mu(y)$ give the transformation from the y coordinates to the y' coordinates if the Jacobian does not vanish i.e. $|\partial y'^\mu / \partial y^\nu| \neq 0$ which we require to be the case. Thus, the operator functions $X'^\mu(X^\mu)$ define a coordinate transformation in $R_n$ between the eigenvalues (coordinates) y and the eigenvalues y' (transformed coordinates) that define $R_n$. Then the set of n real variables $y^\mu$ and the alternative variables $y'^\mu$ both can be interpreted as specifying the coordinates of points in this n-dimensional real space $R_n$ with coordinate transformations given by the functions

$$y'^\mu = y'^\mu(y). \tag{2.12}$$

It now follows that

$$dy'^\mu = (\partial y'^\mu/\partial y^\nu) dy^\nu \tag{2.13}$$

and any set of n functions $V^\mu(y)$ that transform as the coordinates,

$$V'^\mu(y') = (\partial y'^\mu/\partial y^\nu) V^\nu(y) \text{ is to be called a contravariant vector.} \tag{2.14}$$

We use the summation convention for repeated identical indices. The derivatives $\partial/\partial y^\nu$ transform as

$$\partial/\partial y'_\mu = (\partial y^\nu/\partial y'^\mu) \partial/\partial y^\nu \tag{2.15}$$

and any such vector $V_\mu(y)$ which transforms in this manner as

$$V'_\mu(y') = (\partial y^\nu/\partial y'^\mu) V_\nu(y) \text{ is defined as a covariant vector.} \tag{2.16}$$

Upper indices are defined as contravariant indices while lower indices are covariant indices. Functions with multiple upper and lower indices that transform as the contravariant and covariant indices just shown are defined as tensors of the rank of the associated indices.

One would like to have transformations that translate one in the $R_n$ space of the operators X (and thus their eigenvalues y). We define a new additional set of n operators, $D^\mu$, that by definition translates a point an infinitesimal distance, ds, in the $R_n$ respectively in each corresponding direction $y^\mu$ by using the transformation generated by the $D^\mu$ elements of the algebra via the exponential map with transformations:

$$G(ds, \eta) = \exp(ds\, \eta_\mu D^\mu/b) \tag{2.17}$$

In this transformation $\eta_\mu$ is to be a unit vector in the y space, b is an unspecified constant, and ds is defined to be the distance moved in the direction $\eta_\mu$ as defined below. Then

$$X'^\lambda = G X^\lambda G^{-1}. \tag{2.18}$$

By taking ds to be infinitesimal, then one gets

$$X'^\lambda = X^\lambda(s+ds) = \exp(ds\, \eta_\mu D^\mu/b)\; X^\lambda(s)\; \exp(-ds\, \eta_\nu D^\nu/b)$$
$$= (1 + ds\, \eta_\mu D^\mu/b)\; X^\lambda(s)\; (1 - ds\, \eta_\nu D^\nu/b)$$
$$= X^\lambda(s) + ds\, \eta_\mu [D^\mu, X^\lambda]/b + \text{higher order in ds.,} \tag{2.19}$$

Thus the commutator $[D^\mu, X^\lambda]$ defines the way in which the transformations commute (interact) with each other in executing the translations in keeping with the theory of Lie algebras and Lie groups although in general the D & X may not close as a standard Lie algebra. If the space is Euclidian (flat) then there is no dependence of the commutator upon location, and thus there is no interference among the $D^\mu$. Then $[D^\mu, X^\lambda]$ can be normalized to $I \delta_\pm^{\mu\lambda}$ (since $D^\mu$ is defined to translate $X^\mu$) thus

$$[D^\mu, X^\lambda] = I \delta_\pm^{\mu\lambda} = b\, \delta_\pm^{\mu\lambda} \tag{2.20}$$

where $\delta_\pm$ is the diagonal n x n matrix with ±1 on the diagonal with off-diagonal terms zero. This is the customary Heisenberg Lie algebra with structure constants $\delta_\pm^{\mu\lambda}$ and with $[D^\mu, D^\lambda] = 0$ for $\mu \neq \lambda$. The additional operator, I, is to commute with all elements and by definition has a single eigenvalue b, and is needed to close the basis of the Lie algebra which now is of dimension 2n+1. Thus confirming that the distance is ds: $dX^\lambda(s) = ds\, \eta_\mu b/b\, \delta_\pm^{\mu\lambda} = ds\, \eta^\lambda + \text{higher order terms in ds.}$ \tag{2.21}

We now wish to allow for curvature in the space $R_n$ of the X eigenvalues. Thus the [D, X] commutator is now allowed to be dependent upon the operators X and can vary from point to point in the space. We define the functions $g^{\mu\nu}(X)$ as generalized structure functions (no longer constants):

$$[D^\mu, X^\nu] = b\, g^{\mu\nu}(X) \tag{2.22}$$

Where b is a constant to be determined with the requirement that

$$|g| \neq 0 ) \tag{2.23}$$

These generalized structure functions can now also be written as

$$g^{\mu\nu}(X) = [D^\mu, X^\nu]/b \tag{2.24}$$

where $g^{\mu\nu}(X)$ are assumed to be analytic with $g_{\mu\nu}(X)$ defined by

$$g_{\mu\alpha}(X)\, g^{\alpha\nu}(X) = \delta_\mu^{\ \nu} \text{ in the X diagonal representation space.} \tag{2.25}$$

Then using (2.21) one gets

$$X^\mu(s+ds) - X^\mu(s) = dX^\mu = ds\, \eta_\lambda\, g^{\mu\lambda}(X) = ds\, \eta^\mu. \tag{2.26}$$

Then $\quad g_{\mu\nu}(X)\, dX^\mu\, dX^\nu = ds^2\, g_{\mu\nu}(X)\, \eta^\mu\, \eta^\nu = ds^2\,$ since $\eta^\mu$ is a unit vector on this metric, or $\tag{2.27}$

$$ds^2 = g_{\mu\nu}(X)\, dX^\mu\, dX^\nu \text{ proving that } g_{\mu\nu}(X) \text{ is the metric for the space.} \tag{2.28}$$

In the position representation one now has the representation for D as the differential operator:

$$\langle y|\, [D^\mu, X^\nu] = [(b\, g^{\mu\lambda}(y)\, (\partial/\partial y^\lambda) + A^\mu(y),\, y^\nu\,]\, \langle y| = [(\, b\, \partial^\mu + A^\mu(y)\,),\, y^\nu\,]\langle y| = b\, g^{\mu\nu}(y)\, \langle y| \tag{2.29}$$

Thus $\quad \langle y|\, D^\mu = (\, b\, \partial^\mu + A^\mu(y)\,)\, \langle y|$ which allows the D commutator to represent derivatives: $\tag{2.30}$

$$\langle y|\, D^\mu|\Psi\rangle = [b\, g^{\mu\nu}(y)\, ((\partial/\partial y^\nu) + A^\mu(y)\,),\, \Psi(y)] = b\, \partial^\mu\, \Psi(y) \text{ since } [A^\mu(y), \Psi(y)] = 0 \tag{2.31}$$

$$\Psi(y) = \langle y|\Psi\rangle \text{ and } \partial^\mu = g^{\mu\upsilon}(y)\, (\partial/\partial y^\nu) \tag{2.32}$$

and $A^\mu(y)$ is a yet undetermined vector function of $X^\nu$.

The Christoffel symbols are given by

$$\Gamma_{\gamma\alpha\beta} = (\tfrac{1}{2})\, (\partial_\beta\, g_{\gamma\alpha} + \partial_\alpha\, g_{\gamma\beta} - \partial_\gamma\, g_{\alpha\beta}) \tag{2.33}$$

and can be written in the position diagonal representation, in terms of the commutators of D with the metric as

$$\Gamma_{\gamma\alpha\beta} = (\tfrac{1}{2})(1/b)\, (\, [D_\beta, g_{\gamma\alpha}] + [D_\alpha, g_{\gamma\beta}] - [D_\gamma, g_{\alpha\beta}]\,). \tag{2.34}$$

Then using

$$g_{\alpha\beta}(X) = (1/b)\, [D_\alpha, X_\beta] \text{ one obtains} \tag{2.35}$$

$$\Gamma_{\gamma\alpha\beta} = (\tfrac{1}{2})(b^{-2})\, (\, [D_\beta, [D_\gamma, X_\alpha]] + [D_\alpha, [D_\gamma, X_\beta]] - [D_\gamma, [D_\alpha, X_\beta]]\,). \tag{2.36}$$

The Riemann tensor then becomes

$$R_{\lambda\alpha\beta\gamma} = (1/b)\, (\, [D_\beta, \Gamma_{\lambda\alpha\gamma}] - [D_\gamma, \Gamma_{\lambda\alpha\beta}]\,) + (\Gamma_{\lambda\beta\sigma}\, \Gamma^\sigma_{\ \alpha\gamma} - \Gamma_{\lambda\gamma\sigma}\, \Gamma^\sigma_{\ \alpha\beta}) \tag{2.37}$$

where $\Gamma_{\gamma\alpha\beta}$ is to be inserted for the Christoffel symbols giving only commutators. One then defines the Ricci tensor from the Riemann tensor as

$$R_{\alpha\beta} = g^{\mu\nu}\, R_{\alpha\mu\beta\nu} = (1/b)\, [D^\mu, X^\nu]\, R_{\alpha\mu\beta\nu} \text{ and also defines} \tag{2.38}$$

$$R = g^{\alpha\beta}\, R_{\alpha\beta} = (1/b)\, [D^\alpha, X^\beta]\, R_{\alpha\beta}. \tag{2.39}$$

It is well known that the ordinary derivative of a scalar function, $V_\mu = \partial\Lambda/\partial y^\mu$, in Riemann geometry will transform under arbitrary coordinate transformations as a covariant vector. But such a derivative of a vector function of the coordinates will not transform as a tensor. The covariant derivative with respect to $y^\nu$ of a contravariant vector $A^\mu$ is

$$A^\mu_{,\nu} = \partial\, A^\mu/\partial y^\nu + A^\sigma\, \Gamma^\mu_{\ \sigma\nu} \tag{2.40}$$

and the covariant derivative of a covariant vector $A_\mu$ is given by

$$A_{\mu,\nu} = \partial\, A_\mu/\partial y^\nu - A_\sigma\, \Gamma^\sigma_{\ \mu\nu} \tag{2.41}$$

where both $A^\mu_{,\nu}$ and $A_{\mu,\nu}$ transform as tensors with respect to the metric $g^{\alpha\beta}$.

One recalls for Riemannian geometry that there is a Christoffel symbol on the right hand side for each index of the tensor being differentiated. In this algebraic framework one can write the covariant differentiation of a contravariant vector $A^\mu$ as:

$$A^\mu{}_{,\nu} = i [D_\nu, A^\mu] + (\tfrac{1}{2})A^\sigma ( [D_\nu, [D^\mu, X_\sigma]] + [D_\sigma, [D^\mu, X_\nu]] - [D^\mu, [D_\sigma, X_\nu]] ) \tag{2.42}$$

since by definition, A is at most a function of the X operators. Thus we are able to write both the regular derivative (first term) and complete it with the index contraction with the Christoffel symbol (second term). It is important to distinguish this covariant differentiation from the regular differentiation that occurs as a representation of the operator $D^\mu$ in the position representation. It follows that we can write the covariant derivative of any tensor in the same way but with a contraction of the Christoffel symbol with each of the tensor indices as is well known in Riemannian geometry.

One recalls that only the symmetric part of the metric is used to determine distance and angle since it is contracted with a symmetric expression in equation (2.28) as $ds^2 = g_{\mu\nu}(X)\, dX^\mu dX^\nu$. The angle between any two vectors is also defined in the customary way using only the symmetric part of $g_{\mu\nu}(X)$. So one obtains the same geometry when any antisymmetric tensor is added to the metric. There are two natural and important totally antisymmetric tensors which can be used in any linear combination with the metric: (a) the generator for rotations in the $\mu\nu$ plane given by,

$$L^{\mu\nu} = X^\mu D^\nu - X^\nu D^\nu \tag{2.43}$$

which gives the operator for rotations (in physics, the orbital angular momentum tensor) and (b) the antisymmetric tensor

$$F^{\mu\nu} = [D^\mu, D^\nu] \tag{2.44}$$

This allows the generalization of $g^{\mu\nu}$ as $g^{\mu\nu} + a_1 L^{\mu\nu} + a_2 F^{\mu\nu}$ where $a_1$ and $a_2$ are real numbers. (In specific applications these must contain the dimensional components to keep $g^{\mu\nu}$ dimensionless ($\hbar^{-1}$ and $(m_P c)^{-2}$ for applications in physics where $m_P$ is the Plank mass. The $F^{\mu\nu}$ tensor, when one only has the electromagnetic force present, becomes the matrix containing the electric and magnetic fields in physics and thus here is the generalization of those forces due to the intermediate bosons in $A^\mu(X)$ ). While these additional terms do not alter distances and angles, they do introduce changes referred to as torsion in the parallel transport of vectors. Also these changes to $g^{\mu\nu}$ support a gravitational gauge transformation since the geometric properties are invariant under the addition of these antisymmetric components with any values of $a_1$ & $a_2$.

### 3. Application to Physics: The Extended Poincare Algebra & Invariants:

The discussion up to this point is purely mathematical and reframes RG in terms of a GLA. The mathematics is applicable to multiple fields with a non-Euclidian metric, but we now will look at the foundational observable operators in physics when gravitation is not present with QT, SR, and the SM when space-time is Euclidian. The commuting $X^\mu$ operators in the physics setting are operators for events in space-time where the $X^0$ operator has the eigenvalue ct (c = the speed of light and t is time on the reference frame clock while the $X^i$ operators have the eigenvalues of special position, $y^i$, where i ranges over three or more special dimensions. The $D^\mu$ translation operators in the associated dimensions refer to the generalized momentum in the associated direction. The Poincare Lie Algebra (PA) is the fundamental symmetry algebra of physics consisting of the infinitesimal generators of the four (4) translations in space time, $D^\mu$, ($\mu,\nu = 0, 1, 2, 3$), and three (3) rotations & three (3) Lorentz infinitesimal generators with the antisymmetric tensor operator, $M^{\mu\nu}$. All physical theories must be invariant under the Lie group generated by this Lie algebra. As quantum theory is founded upon the relationship between momentum and position operation as defined in the HA, with [D, X] = -iℏ, and [E/c, ct] = [$D^0$, $X^0$] = iℏ, then a full Lie algebra of space-time observables must also include a four-position operator $X^\mu$ in order to formally include the foundations of quantum theory as well as, $M^{\mu\nu}$, the generators of the Lorentz group in a flat space-time. This led us previously to extend the Poincare algebra [20,21] (EPA) by adding a four-vector position operator, $X^\mu$ whose components are to be considered as fundamental observables using a manifestly covariant form of the HA.

We briefly review this EPA algebra whose representations define free particles (fields) in a space-time with no curvature. Our GLA above must reduce to this algebra when there are no forces or gravitation present. As the $X^\mu$ generate translations in momentum, they do not generate symmetry transformations or represent conserved quantities but do provide the critical observables of space-time. We choose the Minkowski metric $g^{\mu\nu}$ = (+1, -1, -1, -1) and write the HA in the covariant form as [$D^\mu$, $X^\nu$] = I $g^{\mu\nu}$ where I is an operator that commutes with all elements and has the unique eigenvalue "iℏ" with $D^0$ = E/c, $D^1$ = $D_x$ …, $X^0$ = ct, $X^1$ = x … where c is the speed of light, E is energy, and t is time. Here the "I" operator is needed to make the fifteen (15) fundamental observables in this extended Poincare (EP) algebra (X, D, M, I) close into a true Lie algebra with the structure constants as follows:

$$[I, D^\mu] = [I, X^\nu] = [I, M^{\mu\nu}] = 0 \tag{3.1}$$

thus I commutes with all operators and has "iℏ" as the only eigenvalue thus:

$$[D^\mu, X^\nu] = i\hbar\, g^{\mu\nu} \tag{3.2}$$

which is the covariant Heisenberg Lie algebra – the foundation of quantum theory,

$$[D^\mu, D^\nu] = 0 \tag{3.3}$$

which insures noninterference of energy momentum measurements in all four dimensions.

$$[X^\mu, X^\nu] = 0 \tag{3.4}$$

which insures noninterference of time and position measurements in all four dimensions.

$$[M^{\mu\nu}, D^\lambda] = i\hbar\, (g^{\lambda\nu} D^\mu - g^{\lambda\mu} D^\nu) \tag{3.5}$$

which guarantees that $D^\mu$ transforms as a vector under $M^{\mu\nu}$ and

$$[M^{\mu\nu}, X^\lambda] = i\hbar\, (g^{\lambda\nu} X^\mu - g^{\lambda\mu} X^\nu)$$

which guarantees that $X^\lambda$ also transforms as a vector under $M^{\mu\nu}$ and thus the Lorentz group is

$$[M^{\mu\nu}, M^{\rho\sigma}] = i\hbar\, (g^{\mu\sigma} M^{\nu\rho} + g^{\nu\rho} M^{\mu\sigma} - g^{\mu\rho} M^{\nu\sigma} - g^{\nu\sigma} M^{\mu\rho}). \tag{3.6}$$

which guarantees that $M^{\rho\sigma}$ transforms as a tensor under the Lorentz group generated by $M^{\mu\nu}$.

The representations of the Lorentz algebra are well-known [8, 9] and are straight forward but the extension to include the four-momentum with the Poincare algebra makes the determination of the Poincare representations somewhat complicated. But with our extension of the Poincare algebra to include a four-position operator, $X^\mu$, the representations are clearer. Because $X^\mu$ is now in the algebra, one can now define the orbital angular momentum four-tensor, operator $L^{\mu\nu}$ as:

$$L^{\mu\nu} = X^\mu D^\nu - X^\nu D^\mu \tag{3.7}$$

From this it can easily be shown that

$$[L^{\mu\nu}, D^\lambda] = i\hbar\, (g^{\lambda\nu} D^\mu - g^{\lambda\mu} D^\nu) \tag{3.8}$$

$$[L^{\mu\nu}, X^\lambda] = i\hbar\, (g^{\lambda\nu} X^\mu - g^{\lambda\mu} X^\nu) \tag{3.9}$$

$$[L^{\mu\nu}, L^{\rho\sigma}] = i\hbar\, (g^{\mu\sigma} L^{\nu\rho} + g^{\nu\rho} L^{\mu\sigma} - g^{\mu\rho} L^{\nu\sigma} - g^{\nu\sigma} L^{\mu\rho}). \tag{3.10}$$

One can then define an intrinsic spin four-tensor as:

$$S^{\mu\nu} = M^{\mu\nu} - L^{\mu\nu} \tag{3.11}$$

with the result that

$$[S^{\mu\nu}, D^\lambda] = 0 ; \quad [S^{\mu\nu}, X^\lambda] = 0 ; \quad [S^{\mu\nu}, L^{\rho\sigma}] = 0; \text{ and} \tag{3.12}$$

$$[S^{\mu\nu}, S^{\rho\sigma}] = i\hbar (g^{\mu\sigma} S^{\nu\rho} + g^{\nu\rho} S^{\mu\sigma} - g^{\mu\rho} S^{\nu\sigma} - g^{\nu\sigma} S^{\mu\rho}) \text{ which is the Lorentz Lie algebra.} \tag{3.13}$$

Now one can separate this EPA algebra into the product of two Lie algebras, the nine parameter HA (consisting of X, P, I) and the six parameter homogeneous Lorentz algebra (consisting of the $S^{\mu\nu}$). Thus one can write all representations as products of the representations of the two algebras. For the HA one can choose the position representation:

$$X^\mu | y > = y^\mu | y > \text{ or the momentum representation} \tag{3.14}$$

$$D^\mu | k > = k^\mu | k > \tag{3.15}$$

or equivalently diagonalize the mass and the sign of the energy and three momenta as

$$D^\mu D_\mu = m^2 c^2; \; \varepsilon(D^0), k^i, \text{ with eigenstates written as } |m, \varepsilon(D^0), k^i > \tag{3.16}$$

The transformations between the position diagonal and momentum diagonal basis vectors are given by the Fourier transform $<y|k> = (2\pi)^{-2} \exp((g_{\mu\nu} y^\mu k^\nu)/(i\hbar))$ as is well known.

All representations of the homogeneous Lorentz group have been found by Bergmann and by Gelfand, Naimark, and Shapiro [2,9] to be given by the two Casimir operators $b_0$ and $b_1$ defined as:

$$b_0^2 + b_1^2 - 1 = \tfrac{1}{2} g_{\mu\rho} g_{\nu\sigma} S^{\mu\nu} S^{\rho\sigma} \tag{3.17}$$

where $b_0 = 0, \tfrac{1}{2}, 1, 3/2, ...(|b_1|-1)$ and where $b_1$ is a complex number defined by

$$b_0 b_1 = -\tfrac{1}{4} \varepsilon_{\mu\nu\rho\sigma} S^{\mu\nu} S^{\rho\sigma} \tag{3.18}$$

with the rotation Casimir operator as $S^2$ which has the spectrum $s(s+1)$ with the total spin

$$s = b_0, b_0+1, ..., (|b_1| - 1) \tag{3.19}$$

and the z component of spin:

$$\sigma = -s, -s+1, ....s-1, s \tag{3.20}$$

Thus the homogeneous Lorentz algebra representation can be written as $| b_0, b_1, s, \sigma >$ which joined with the Heisenberg algebra gives the full representation space as either

$$|k^\mu, b_0, b_1, s, \sigma > = a^+_{k, b_0, b_1, s, \sigma} |0> \text{ for the momentum representation or} \tag{3.21}$$

$$| y^\mu, b_0, b_1, s, \sigma > = a^+_{y, b_0, b_1, s, \sigma} |0> \text{ for the position representation.} \tag{3.22}$$

To obtain the effective extended Poincare algebra, one must generalize the equations above by allowing the effective momentum, $D^\mu$, to contain both the gravitational metric and the vector bosons A as in equation (2.29) and likewise write

$$L^{\mu\nu} = X^\mu D^\nu - X^\nu D^\mu. \tag{3.23}$$

The EPA representations expressed as creation and annihilation operations (to allow for particle creation and destruction) are then used to frame all expressions with the introduction here of the metric as a spin two (graviton) particle with a zero mass. As one can see, free particles are to be represented by the representations of the EPA shown above. But as mentioned above, these representations give an infinite number of diverse spin representations and continuous mass representations – even imaginary mass. There is no constraint between spin and mass as that which is found in the real world. This is where the phenomenological SM (SU(3) x SU(2) x U(1)) gauge Lie algebra comes to the rescue by only allowing certain spins and certain masses and certain interactions. But the SM comes from observations and not from fundamental principles which is what one desires.

<u>Invariance of the speed of light with $X^0$ as ct:</u> If the X operators are to refer to physical space-time, and the space is Euclidian, then the theory of special relativity requires that the transformation from one frame to another preserve the invariance of the speed of light $ds^2 = g_{\mu\nu} dX^\mu dX^\nu = g'_{\mu\nu} dX'^\mu dX'^\nu$ or less formally stated as $ds^2 = c^2t^2 - r^2$ is invariant which introduces the speed of light, c, into the equations.

<u>Connection with the Heisenberg Algebra and QT (with ℏ)</u>: One notes that equation [2.22] must reduce to the Heisenberg algebra in a flat space-time when $g^{\mu\nu}(X) = g^{\mu\nu}$ and does not depend upon the position operators. To get

$$[D^\mu, X^\nu] = i\hbar\, \delta_\pm^{\mu\lambda} = i\hbar\, g^{\mu\nu} \tag{3.24}$$

<u>one must set the previous constant b = iℏ</u> which introduces Planks constant into the equations and determines "b" so that the system reduces to the standard HA for flat space-time with X representing the operators for space-time and D as the effective momentum of a particle. Thus, in a flat space-time one retrieves the standard HA and the associated QT. It is of interest to see that <u>the metric of curved space-time is the measure of the interference between the effective momentum and position in space time.</u> Equation (3.2) now introduces and in fact automatically will force the factor iℏ into Einstein's equations for GR.

<u>Connection with Special Relativity and the Lorentz and Poincare Algebra (c):</u>

The connection with special relativity and the Lorentz algebra, follows from the interpretation of the four-position operators $X^\mu$ and their eigenvalues as $X^0 = ct$, and $X^i = (x, y, z)$ with i = 1, 2, & 3 where c is the speed of light with SI units of meters per seconds and $D^\mu$ as the effective four-momentum with eigenvalues (E/c, $P^i$). In a theory with additional dimensions, subsequent values of $X^i$ would refer to hidden dimensions. In this limit one gets

$$g_{\alpha\beta} D^\alpha D^\beta = m^2 c^2 \tag{3.25}$$

giving the mass of the particle, and

$$\langle y | g_{\alpha\beta}\, dX^\alpha\, dX^\beta = g_{\alpha\beta}\, dy^\alpha\, dy^\beta \langle y | = \langle y | d\tau^2 \tag{3.26}$$

giving the proper time (invariant infinitesimal path length of elapsed time on the particle). This then gives the invariance of the speed of light from $d\tau^2 = c^2 dt^2 - dr^2$ among inertial frames. But we note that additional dimensions are not ruled out allowing for other extensions to this theory.

<u>Connection with the SM</u>: The $A^\mu$ field arose naturally as part of the general solution to the equation $[D^\mu, X^\nu] = i\hbar\, g^{\mu\nu}(y)$ in the position diagonal representation. This $A^\mu$ vector field can contain any spectrum of masses but only a select few are seen in nature. The standard model of particle physics specifies that the gauge transformations are representations of the U(1) x SU(2) x SU(3) algebras along with an associated model which specifies the masses and all interactions with the exception of gravitation. So $\langle y | D^\mu = (b\, g^{\mu\nu}(y)\, \partial/\partial y^\mu + A^\mu(y))\langle y |$ gives the gauge covariant form for the translation operator $D^\mu$. But if this operator act upon a vector or tensor with indices, then there is another term that must be included in order for the derivative of a non-scalar field to transform as a vector in the Riemann space: $\langle y | D^\mu = (b\, g^{\mu\nu}(y)\, \partial/\partial y^\mu + \Gamma(y) + A^\mu(y))\langle y |$ where Riemann covariance requires the actual form: $A^\mu{}_{,\nu} = \partial A^\mu / \partial y^\nu + A^\sigma \Gamma^\mu{}_{\sigma\nu}$ as discussed above for action on a vector or tensor field with the Christoffel symbol. So not only does $D^\mu$ contain the fields for all forces in the SM, it also always contains the gravitational field $g^{\mu\nu}(y)$ and when acting on any vector or tensor also contains the gravitational field via the Christoffel and $g^{\mu\nu}(y)$ terms. So, $D^\mu$ contains all gauge particles (forces) representing the intermediate bosons. One notes that there were no additional assumptions made and this is a natural outcome of proceeding from the expression of Riemannian geometry as a generalized Lie algebra.

### Einstein's equations for GR:

We have already done the necessary background work to frame GR in terms of the operator algebra. In general relativity the Einstein equations for the metric:

$$R_{\alpha\beta} - \tfrac{1}{2} g_{\alpha\beta} R + g_{\alpha\beta}\Lambda = (8 \pi G/c^4) \, T_{\alpha\beta} \quad \text{can now be written} \tag{3.27}$$

$$R_{\alpha\beta} - (i\hbar \,[D_{\alpha}, X_{\beta}]) \, ( \tfrac{1}{2} R - \Lambda ) = (8 \pi G/c^4) \, T_{\alpha\beta} \tag{3.28}$$

where $R_{\alpha\beta}$ and $R$ are now given in terms of commutators as shown above in (2.38) and (2.39) while $T_{\alpha\beta}$ is the energy-momentum tensor as determined by the SM. Thus, all terms on the LHS consist only of commutators and (3.9) with the Einstein equations in GR expressed totally as GLA thus framing GR as a non-commutative algebra of the X, D and g(X) observables and their eigenvalues.

## 4. Implications of this Model:

(1.) <u>Planks Constant in GR:</u>  The first consequence of our generalized algebraic formulation is that Einstein's equations for GR now MUST contain Planks constant (iℏ) in the exact way shown and which is also now present in the R and $R_{\alpha\beta}$ terms. Previously, GR was framed in a purely classical manner and the exact method of quantization of GR was not clear for incorporation with the standard model. In this model with the GLA, the introduction of Planks constant is unambiguous and mandatory. In fact, the Heisenberg Lie algebra follows from the identification of $D^{\mu}$, the generalized momentum, as the translation operator for the space coordinates $X^{\mu}$.

(2.)  One of the $X^{\mu}$ variables ($X^0$) must refer to time (as a constant "c" times "t" while the others refer to the location $y^i$ in space of 3 (or n) dimensions as the eigenvalues of these simultaneous commuting observables for the location of events. The invariance of c in all inertial reference frames implies that the transformation group among frames is given by the Poincare group and is the foundation of SR.

(3.)  We know that in physics, the (covariant) divergence of the energy momentum tensor vanishes in closed systems which gives us the conservation of the energy momentum four-vector in closed systems. We also know that the energy and momentum of the particles that exist are the only possible source of any curvature of the space time. It follows then, as Einstein originally argued, that since the (covariant) divergence of the Einstein tensor also vanishes then the Einstein tensor must be proportional to the energy momentum tensor. The determination of that proportionality then gives the Einstein equations for the curvature of space-time as we have written in algebraic commutator form above.

(4.) <u>Gauge Transformations:</u> One can generalize the gauge transformation

$$|\Psi\rangle \rightarrow e^{\Lambda} |\Psi\rangle \quad \text{to} \quad |\Psi\rangle \rightarrow e^{\Lambda(x)} |\Psi\rangle = |\Psi'\rangle \tag{4.1}$$

since this leaves invariant the scalar product $\langle\Psi|\Psi\rangle$ invariant.
But a full invariance requires that the action of $D^{\mu}$ on $|\Psi'\rangle$ must also leave the scalar product invariant. Then it follows that when

$$\langle y|D^{\mu} = \langle y| \, ( i\hbar \, g^{\mu\beta}(y) \, \partial/y^{\beta} + A^{\mu}(y) ) \text{ acts upon } e^{\Lambda(x)} |\Psi\rangle \text{ one gets:} \tag{4.2}$$

$$D^{\alpha} \langle y|(e^{\Lambda(x)} |\Psi\rangle) = ( i\hbar \, g^{\mu\beta}(y) \, \partial/y^{\beta} + A^{\mu}(y) ) \, (e^{\Lambda(y)} \langle y||\Psi\rangle) \tag{4.3}$$

$$= e^{\Lambda(y)} ((i\hbar \, g^{\mu\beta}(y) \, \Lambda(y) /y^{\beta} ) + ( i\hbar \, g^{\mu\beta}(y) \, \partial/y^{\beta} + A^{\mu}(y) ) \, (\langle y||\Psi\rangle)). \tag{4.4}$$

So, it follows that A must also transform as

$$A^{\mu}(y) \rightarrow A^{\mu}(y) - (i\hbar \, g^{\mu\beta}(y) \, \Lambda(y) /y^{\beta} ) \tag{4.5}$$

to cancel the derivative of the position dependent phase $\Lambda(y)$.

Generally, one can argue that that which exists, $|\Psi>$, must be a representation space of the algebra of operators X and D which in empty space have the commutation rules of the extended Poincare algebra of $X^\mu$, $D^\mu$, and $M^{\mu\nu}$. Those representations are known but they allow all spins and masses representing the invariants of that algebra. But that is not what we find in nature rather there are a collection of masses and associated spins that closely adhere to the standard model of particle physics: U(1) x SU(2) x SU(3) with multiple seemingly arbitrary parameters. This "Standard Model" has evolved over the last few decades with incredible success, but we do not have a foundational theory for it and do not have a way to determine the associated masses, spins, and associated parameters. But such quantization of the associated Casimir operations of the XPM algebra when P is replaced by $D^\mu$, $M^{\mu\nu}$ by $L^{\mu\nu} + S^{\mu\nu}$ where $L^{\mu\nu} = X^\mu D^\nu - X^\nu D^\mu$. Such representations are of great difficulty, but it is possible that they could lead us to certain quantized states that would shed light on the basis of the standard model.

(5.) <u>Altered Uncertainty Principle:</u> The expression of Einstein's equations in the form of commutators of operators leads to a third result: In a strong gravitational field near a star, such as a non-rotating white dwarf, one can treat the metric as constant using the Schwarzschild solution over a region that is small relative to the size of the star. The radial direction can be taken as the $y^1$ direction as the distance to the center of the star, with

$$g_{00} = (1 - r_s/y^1) \text{ and } g_{11} = -1/(1 - r_s/y^1) \tag{4.6}$$
$$\text{where } r_s = 2GM/c^2 \text{ with } g_{22} = g_{33} = -1 \tag{4.7}$$

and where G is the gravitational constant, M is the mass of the star, c is the speed of light, and $y^1$ is the distance to the center of the star giving g(X) as the Schwarzschild solution. Equation (3.8) is still exactly the classical equation for the metric but recast in commutation relations (3.9). We therefore require that equation (3.9) be satisfied to determine the metric. However, there are certain results that follow that could potentially test our approach: The position and momentum operators are now to have the interpretational structure given by quantum theory with free particles as representations of the position extended Poincare algebra. The essentially new feature is that by virtue of the presence of a particle in a gravitational field such as near a star, the commutation rules with the rationalized Planks constant, ℏ, are effectively modified by the metric in the radial ($X^1$) and time ($X^0$) directions with the specific prediction that in a small region, with the Schwarzschild metric, one gets the altered uncertainty principles:

$$\Delta X_r \, \Delta D_r \geq (\hbar/2)(1/(1-r_s/r)) \text{ and} \tag{4.8}$$
$$\Delta t \, \Delta E_r \geq (\hbar/2)(1-r_s/r) \tag{4.9}$$

where $r_s = 2GM/c^2$ and where r = the distance to the center of the spherical mass. This is because the generalized algebra effectively alters the value of Planks constant in both the $X^1$ and $X^0$ directions as well as the wave nature of particles in the altered local Fourier transform. <u>This would in turn alter the creation rate of virtual pairs in the vacuum in a strong gravitational field. What is maintained is a more general form of the Heisenberg uncertainty principle</u> obtained by multiplying (3.33) and (3.34) together to obtain

$$\Delta t \, \Delta E_r \, \Delta X_r \, \Delta D_r \geq (\hbar/2)^2 \tag{4.10}$$

while the other two uncertainty relations remain unchanged. The metric would be quantized along with the vector fields in D.

## 4. Conclusions:

a. The first and primary result of this work is the purely mathematical generalization of a Lie algebra composed of an Abelian algebra of n operators $X^\nu$, and n translation operators in that space, $D^\mu$, that have the commutation relations $[D^\mu, X^\nu] = I\, g^{\mu\nu}(X)$ as described, where the functions $g^{\mu\nu}(X)$ generalize the previously constant "structure constants" of a Lie algebra to allow them to be functions of the Abelian subalgebra where I commutes with all elements and has a constant eigenvalue "b". We showed that within this framework that <u>Riemannian geometry can be recast as a non-commutative algebra (NCA) with $D^\mu$ represented as $D^\mu = b\,(\,g^{\mu\upsilon}(y)\,(\partial/\partial y^\upsilon) + \Gamma + A^\mu(y)\,)$ on the X diagonal representation space and that the structure of Riemannian geometry can be fully expressed in such commutation relations</u>.

      b. Furthermore, it followed that $ds^2 = g_{\mu\nu}(X)\,dX^\mu\,dX^\nu$ thus $g_{\mu\nu}(X)$ was the metric of that space which reduces to the Heisenberg algebra metric $\delta_\pm^{\mu\lambda}$ when the space is not curved. Thus, <u>the metric $g_{\mu\nu}(X)$ captures the interference in the commutators for a curved space</u>.

      c. Then when this framework is applied to fundamental quantum mechanics, <u>the constant "b" must reduce to the Plank constant multiplied by "i"</u>.

      d. This in turn <u>forces</u> the unique introduction of $i\hbar$ into the Einstein equations when they are expressed in terms of this commutation relations.

      e. The conserved effective momentum <u>$D^\mu = i\hbar\,(\,g^{\mu\upsilon}(y)\,(\partial/\partial y^\upsilon) + \Gamma + A^\mu(y)\,)$ both allows the intermediate bosons of the SM and combines it with the metric which must be quantized as a spin two field. This then integrates the metric (graviton) with the other boson fields explicitly for local gauge transformations which are now supported in this framework</u>.

      f. It follows from the generalized commutation rules, $[D^\mu, X^\nu] = I\,g^{\mu\nu}(X)$, that <u>the uncertainty relations between energy and time, and between momentum and space are altered in the presence of gravitational fields as $\Delta X^i\,\Delta D^j \geq (\hbar/2)\,g^{ij}(y)$ and $\Delta t\,\Delta E_r \geq (\hbar/2)\,g^{00}(y)$</u>.

      g. The <u>generator for rotations in the $\mu\nu$ plane is given by $L^{\mu\nu} = X^\mu\,D^\nu - X^\nu\,D^\nu$</u> which gives the the orbital angular momentum tensor in this system. The antisymmetric tensor <u>$F^{\mu\nu} = [D^\mu, D^\nu]$ represents the associated forces of the standard model</u> with the (rather complicated) presence of the metric.

      h. In keeping with relativistic quantum theory, particles (fields) are to be <u>represented by creation and annihilation operators, indexed by representations of the (extended) Poincare algebra.</u> It is notable that there is very little freedom in this model except that the space time can be of any dimension and the system is subject to the current SM with the metric being determined by the energy momentum tensor in keeping with Einstein's equations.

## 5. The Asymmetric Components of the Metric – Next Steps:

      The analysis above deals only with the 10 symmetric components of the metric $g^{\mu\nu}{}_{(s)}$ and the means of including the standard model in the effective momentum $D^\mu$. These components, determine both length and angle and thus the associated geometry as given by Einstein's equations that link the (symmetric) energy-momentum tensor density $T^{\mu\nu}$ to the Einstein tensor $G^{\mu\nu}$, both of which have zero covariant divergence, thus determining the symmetric components of $g^{\mu\nu}{}_{(s)}$. But there are 6 more components of the metric that are left undetermined that can be associated with other aspects of the space-time structure. We will next investigate the use of the angular momentum tensor $M^{\mu\nu} = X^\mu\,D^\nu - X^\nu\,D^\mu$ and the angular momentum tensor density $m^{\mu\nu\rho} = X^\mu\,T^{\nu\rho} - X^\nu\,T^{\mu\rho}$ which has zero covariant divergence as potentially determining the 6 asymmetric components $g^{\mu\nu}{}_{(a)}$ as determined by the conjectured equation **$X^\mu\,G^{\nu\rho} - X^\nu\,G^{\mu\rho} = \beta\,m^{\mu\nu\rho}$** where both sides have zero covariant divergence and $\beta$ is an undetermined constant. Dark matter initially was identified from the aberrations of rotational motions in galaxies and perhaps this approach could lead to a complementary set of equations to those of Einstein that could shed light on these rotational anomalies attributed to dark matter using angular momentum densities as a source instead of the energy-momentum tensor density.

lly but is imposed by the phenomenological SM.